\documentclass[12pt]{article}
\usepackage{graphicx}
\usepackage{amsmath,amssymb}
\usepackage{bm}
\usepackage{graphicx, color}
\usepackage{wrapfig}
\begin{document}
\title{
\begin{flushright}
\ \\*[-80pt] 
\begin{minipage}{0.2\linewidth}
\normalsize
\end{minipage}
\end{flushright}
{\Large \bf  Direct CP violation of $b \rightarrow s\gamma$ 
and  CP asymmetries
 of non-leptonic  $B$ decays  
 in  squark flavor mixing
\\*[20pt]}}

\author{
\centerline{Yusuke~Shimizu$^{1,}$\footnote{E-mail address: shimizu@muse.sc.niigata-u.ac.jp}, \ \ 
Morimitsu~Tanimoto$^{1,}$\footnote{E-mail address: tanimoto@muse.sc.niigata-u.ac.jp}, \ \  and \ \
Kei~Yamamoto$^{2,}$\footnote{E-mail address: yamamoto@muse.sc.niigata-u.ac.jp}}
\\*[20pt]
\centerline{
\begin{minipage}{\linewidth}
\begin{center}
$^1${\it \normalsize
Department of Physics, Niigata University,~Niigata 950-2181, Japan }
\\*[4pt]
$^2${\it \normalsize
Graduate~School~of~Science~and~Technology,~Niigata~University, \\ 
Niigata~950-2181,~Japan }
\end{center}
\end{minipage}}
\\*[70pt]}

\date{
\centerline{\small \bf Abstract}
\begin{minipage}{0.9\linewidth}
\vskip  1 cm
\small
 We study the contribution of the squark flavor mixing 
from  the $LR(RL)$ component of the squark mass matrices
to the direct CP violation of the $b\to s\gamma$ decay and
the CP-violating asymmetry in the non-leptonic decays of 
$B$ mesons.
The magnitude of the  $LR(RL)$ component is constrained
by the branching ratio and the direct CP violation  of $b\to s\gamma$.
We predict  the correlation of the CP asymmetries
among  $A_{\text{CP}}^{b \to s \gamma}$, $\mathcal{S}_{\phi K_S}$
and $\mathcal{S}_{\eta 'K^0}$ of the $B$ decays.
 The precise data of these  CP violations will give us the crucial test
 for our  framework of the squark flavor mixing.
\end{minipage}
}

\begin{titlepage}
\maketitle
\thispagestyle{empty}
\end{titlepage}

\section{Introduction}
\label{sec:Intro}

New physics are expected to be observed at the  LHC experiments.
Although new particles have not yet discovered, LHCb has  reported
 new data of the CP violation of $B$ mesons and the branching ratios
 of rare $B$ decays.  New physics are  also expected to be
 found in the $B$ meson  decays.

The CP violation in the $K$ and $B_d$ mesons has been successfully
understood 
within the framework of the standard model (SM),
 so called Kobayashi-Maskawa (KM) model \cite{Kobayashi:1973fv}.
The source of the CP violation is the KM phase 
in the quark sector with three families. 
However, there could be new sources of the CP violation if the SM is 
extended to the supersymmetric (SUSY) models. The CP-violating phases appear
in  soft scalar mass matrices. These phases  contribute to flavor changing 
neutral  currents  with the CP violation.
Therefore, we expect the SUSY contribution in the CP-violating phenomena
in the $B$ meson decays.
 
The typical contribution of SUSY is 
the  gluino-squark mediated flavor changing process
 \cite{King:2010np}-\cite{Ishimori:2011nv}. 
In our previous paper~\cite{Hayakawa:2012ua}, 
we have already discussed  the effect of the squark flavor mixing 
on the CP violation in the non-leptonic decays
of $B_d^0$ and $B_s^0$
taking  account of the recent LHCb experimental data.
 We have found the deviation from the SM predictions
 in  the asymmetries of  the penguin dominated decays 
 $B_d^0\to \phi K_S$ and $B_d^0\to \eta 'K^0$.
In that framework of the SUSY contribution, we assume that
 $LR$ and $RL$ components of the squark mass matrices are neglected.
 The $LL$ and $RR$ components of squark mass matrices  
contribute considerably to the penguin  processes
 for the case of  large $\mu\tan\beta$, ${\cal O}(10~{\rm TeV})$.
However,  if $LR$ and $RL$ components of squark mass matrices
  dominate the penguin decays, 
 the asymmetries of 
 $B_d^0\to \phi K_S$ and $B_d^0\to \eta 'K^0$ are deviated from the
SM predictions even for  the case of 
the smaller  $\mu\tan\beta$, ${\cal O}(1~{\rm TeV})$.
Then, these contributions  of the new physics 
 are  correlated with the direct CP violation  of the  $b\to s\gamma$ decay.
In this paper, we present the numerical analyses in the  case that 
$LR$ and $RL$ components of squark mass matrices dominate the penguin decays.
In this case, the $LR(RL)$ components do not contribute to 
 the  dispersive part $M_{12}^q$ of $B_q-\bar B_q \ (q=d,s)$ mixing.

In section 2,
we summarize the effect of  new physics in the CP violations 
 of the neutral $B$ mesons
  including the recent  experimental  data.
In section 3,
we discuss the our framework  of  the squark flavor mixing in
the CP violation of $B$ mesons.
  We also discuss the constraints from the direct CP violation
 in  the $b\to s\gamma$ process.
 In section 4, we show the numerical result of 
the CP violation in the  $B$ mesons.
Section 5 is devoted to the summary.

\section{New physics of  CP violation in  $B$ mesons}
\label{sec:Deviation}


Let us discuss the effect of  new physics 
 in the non-leptonic decays of $B$ mesons. 
The contribution of new physics to the dispersive part $M_{12}^q(q=d,s)$ 
is parameterized as 
\begin{equation}
M_{12}^q=M_{12}^{q,\text{SM}}+M_{12}^{q,\text{SUSY}}=
M_{12}^{q,\text{SM}}(1+h_qe^{2i\sigma _q})~, \quad (q=d,s)
\label{M12}
\end{equation}
where $M_{12}^{q,\text{SUSY}}$ is the SUSY contribution, 
and the SM contribution $M_{12}^{q,\text{SM}}$ is given as \cite{sanda}
\begin{equation}
M_{12}^{q,\text{SM}}=\frac{G_F^2M_{B_q}}{12\pi ^2}M_W^2(V_{tb}V_{tq}^*)^2\hat \eta _BS_0(x_t)f_{B_q}^2B_q~.
\end{equation} 

The time dependent  CP asymmetry decaying into the final state $f$
 is defined as \cite{Aushev:2010bq}
\begin{equation}
\mathcal{S}_{f}=\frac{2\text{Im}\lambda _{f}}{|\lambda_{f}|^2+1}\ ,
\label{sf}
\end{equation}
where 
\begin{equation}
\lambda_{f}=\frac{q}{p} \bar \rho\ , \qquad 
\frac{q}{p}=\sqrt{\frac{M_{12}^{q*}-\frac{i}{2}\Gamma_{12}^*}
{M_{12}^{q}-\frac{i}{2}\Gamma_{12}}}, \qquad
\bar \rho \equiv
\frac{\bar A(\bar B_q^0\to f)}{A(B_q^0\to f)}.
\end{equation}
In the  $B_d^0\to J/\psi  K_S$ decay, $\lambda_{J/\psi  K_S}$
is given in terms of  the new physics parameters
$h_d$ and $\sigma_d$ as
\begin{equation}
\lambda_{J/\psi  K_S}=
-e^{-i\phi _d},\quad \phi _d
=2\beta_d+\text{arg}(1+h_de^{2i\sigma _d}),
\label{new}
\end{equation}
by putting  $|\bar \rho |=1$ and $q/p\simeq \sqrt{M_{12}^{q*}/M_{12}^q}$,
where the phase $\beta_d $ is given  in the SM.
The CKMfitter provided the allowed region of $h_d$ and $\sigma_d$,
where the central values are \cite{CKMfitter,Ligeti}
\begin{equation}
h_d\simeq 0.3,\qquad \sigma _d\simeq 1.8 \ {\rm rad}.
 \label{hdsigmaLHCb}
\end{equation}
 Since  penguin processes are dominant  in  the case of
  $f=\phi K_S, \eta ' K^0$, the loop induced new physics
 could  contribute considerably 
on the   CP violation of  those  decays.
 Then,  $\mathcal{S}_f$ is not any more same 
as $\mathcal{S}_{J/\psi K_S}$
 if new physics leads to  $|\bar \rho |\not = 1$.
 Those predictions provide us good tests for  new physics.

In the   $B_s^0\to J/\psi\phi$ decay, we parametrize as
\begin{equation}
\lambda _{J/\psi \phi }=
e^{-i\phi _s},\qquad \phi _s
=-2\beta_s+\text{arg}(1+h_se^{2i\sigma _s}),
\end{equation}
where $\beta _s$ is given  in the SM.
Recently the  LHCb presented the observed 
 CP-violating phase $\phi _s$ in $\bar B_s^0\to J/\psi \pi^+\pi^- $ 
decay using about 1 fb$^{-1}$ of data~\cite{:2012dg}.
This result leads to
\begin{equation}
\phi _s=-0.019^{+0.173+0.04}_{-0.174-0.03} \ \ \text{rad},
\label{phis}
\end{equation}
which is consistent with the SM prediction \cite{CKMfitter}
\begin{equation}
\phi _s^{J/\psi \phi ,SM}=-2\beta_s=-0.0363\pm 0.0017~\text{rad}.
\label{SMBs}
\end{equation}
Taking  account of these data, the CKMfitter has presented
the allowed values of $h_s$ and $\sigma _s$ ~\cite{CKMfitter,Ligeti}.
The allowed region is rather large including zero values.
In order to investigate possible contribution of  new physics,  we take 
the central values
\begin{equation}
h_s= 0.1,\qquad \sigma _s=0.9 - 2.2 \ \text{rad},
\label{hsigmaLHCb}
\end{equation}
as a typical parameter set in our work.

We remark on numerical inputs of phases $\phi_d$ and $\phi_s$
in our calculation.
The phase  $\phi_d$ is derived  from  the
 observed value  $\mathcal{S}_f=0.671\pm 0.023$ in $B_d^0\to J/\psi K_S$
\cite{PDG} as seen  Eqs.(\ref{sf}) and (\ref{new}).
On the other hand, we use the SM value of $\beta_s$ and
 the values of the new physics  parameters,  $h_s$ and $\sigma_s$ 
in Eq.(\ref{hsigmaLHCb})
 to estimate  $\phi_s=-2\beta_s+\text{arg}(1+h_s e^{2i\sigma_s})$. 
 We do not use the observed value of $\phi_s$ in $B_s^0\to J/\psi\phi$
 because of  the large experimental error in Eq.(\ref{phis}).

Since the $B_d^0\to J/\psi K_S$ process occurs at the tree level in  SM, 
 the CP-violating  asymmetry  originates   from  $M_{12}^d$.
Although the $B_d^0\to \phi K_S$ and $B_d^0\to\eta 'K^0$ decays 
are penguin dominant ones,
their asymmetries also come from  $M_{12}^d$.
Then,  asymmetries of
 $B_d^0\to J/\psi K_S$,  $B_d^0\to \phi K_S$ and 
$B_d^0\to \eta 'K^0$ are expected to be same magnitude in SM.

On the other hand, 
 if the squark flavor mixing  contributes to the decay 
 at the one-loop level, its magnitude could be  comparable 
to the SM penguin one
 in  $B_d^0\to \phi K_S$ and $B_d^0\to \eta 'K^0$, 
but it is tiny in $B_d^0\to J/\psi K_S$. 
Endo, Mishima and Yamaguchi proposed the possibility to
 find the SUSY contribution  in these asymmetries 
\cite{Endo:2004dc}.
The  present data suggest the deviation from SM
in these time dependent asymmetries  of  $B_d^0$ decays such as,
\begin{equation}
\mathcal{S}_{ J/\psi K_S}=0.671\pm 0.023, \qquad
\mathcal{S}_{\phi K_S}=0.39\pm 0.17, \qquad  
\mathcal{S}_{\eta 'K^0}=0.60\pm 0.07 ,
\label{Sfdata}
\end{equation}
 however, precise data are required to justify the new physics contribution.

New physics contribute to the $b\to s\gamma$ process.
The observed $b\to s\gamma$ branching ratio (BR) is 
$(3.60\pm  0.23)\times 10^{-4}$ \cite{PDG}, on the other hand 
the SM prediction is given as $(3.15\pm  0.23)\times 10^{-4}$
at ${\cal{O}}(\alpha_s^2)$
\cite{Buras:1998raa,Misiak:2006zs}.
Therefore, the contribution of  new physics should be suppressed. 
New  physics are  also constrained
 by the direct CP violation  
\begin{equation}
A_{\text{CP}}^{b\to s\gamma}
\equiv
\frac{\Gamma (\bar{B} \to X_{s}\gamma )
-\Gamma (B \to X_{\bar{s}}\gamma )}
{\Gamma (\bar{B} \to X_{s}\gamma )
+\Gamma (B \to X_{\bar{s}}\gamma )} .
\end{equation}
Since SM prediction   $A_{\text{CP}}^{b\to s\gamma}\simeq 0.005$
is tiny \cite{Kagan:1998bh}, new physics may appear in this CP asymmetry.
The present data $A_{\text{CP}}^{b\to s\gamma}=-0.008\pm 0.029$ \cite{PDG}
has rather 
large error bar, and so the constraint of  new physics is not so severe.
However, improved data will provide the crucial test for  new physics.
\section{Squark flavor mixing and CP violations of $B$ mesons}
\label{sec:Squark}
Let us consider the flavor structure of squarks
in order to estimate the CP-violating asymmetries of $B$ meson decays.
We take the most popular anzatz, which is to  postulate 
a degenerate SUSY breaking mass spectrum for down-type squarks.
Then, in the super-CKM basis, we can parametrize 
 the soft scalar masses squared 
$M^2_{\tilde d_{LL}}$, $M^2_{\tilde d_{RR}}$, 
$M^2_{\tilde d_{LR}}$, and $M^2_{\tilde d_{RL}}$ for the down-type squarks
as follows:
\begin{align}
M^2_{\tilde d_{LL}}&=m_{\tilde q}^2
\begin{pmatrix}
1+(\delta _d^{LL})_{11} & (\delta _d^{LL})_{12} & (\delta _d^{LL})_{13} \\
(\delta _d^{LL})_{12}^* & 1+(\delta _d^{LL})_{22} & (\delta _d^{LL})_{23} \\
(\delta _d^{LL})_{13}^* & (\delta _d^{LL})_{23}^* & 1+(\delta _d^{LL})_{33}
\end{pmatrix}, \nonumber \\
M^2_{\tilde d_{RR}}&=m_{\tilde q}^2
\begin{pmatrix}
1+(\delta _d^{RR})_{11} & (\delta _d^{RR})_{12} & (\delta _d^{RR})_{13} \\
(\delta _d^{RR})_{12}^* & 1+(\delta _d^{RR})_{22} & (\delta _d^{RR})_{23} \\
(\delta _d^{RR})_{13}^* & (\delta _d^{RR})_{23}^* & 1+(\delta _d^{RR})_{33}
\end{pmatrix}, \nonumber \\
M^2_{\tilde d_{LR}}&=(M_{\tilde d_{RL}}^2)^\dagger =m_{\tilde q}^2
\begin{pmatrix}
(\delta _d^{LR})_{11} & (\delta _d^{LR})_{12} & (\delta _d^{LR})_{13} \\
(\delta _d^{LR})_{21} & (\delta _d^{LR})_{22} & (\delta _d^{LR})_{23} \\
(\delta _d^{LR})_{31} & (\delta _d^{LR})_{32} & (\delta _d^{LR})_{33}
\end{pmatrix},
\end{align}
where  $m_{\tilde q}$ is the average squark mass, and
$(\delta _d^{LL})_{ij}$, $(\delta _d^{LR})_{ij}$, $(\delta _d^{RL})_{ij}$, 
and $(\delta _d^{RR})_{ij}$ are  called as
the  mass insertion (MI) parameters.
The MI parameters are supposed to be much smaller than $1$.

The contribution of the gluino-squark box diagram to the dispersive part of the effective Hamiltonian for 
the $B_q$-$\bar B_q$ mixing is written as \cite{Gabbiani:1996hi,Altmannshofer:2009ne} 
\begin{align}
M_{12}^{q,SUSY}&=A_1^q\Big [A_2\left \{ (\delta _d^{LL})_{ij}^2+
(\delta _d^{RR})_{ij}^2\right \} +
A_3^q(\delta _d^{LL})_{ij}(\delta _d^{RR})_{ij} \nonumber \\
&+A_4^q\left \{ (\delta _d^{LR})_{ij}^2+(\delta _d^{RL})_{ij}^2\right \} +A_5^q(\delta _d^{LR})_{ij}(\delta _d^{RL})_{ij}\Big ],
\label{bbbarmixing}
\end{align}
where 
\begin{align}
&A_1^q=-\frac{\alpha _S^2}{216m_{\tilde q}^2}\frac{2}{3}M_{B_q}f_{B_q}^2,
\qquad A_2=24xf_6(x)+66\tilde f_6(x),\nonumber \\
&A_3^q=\left \{ 384\left (\frac{M_{B_q}}{m_j+m_i}\right )^2+72 \right \} 
xf_6(x)+\left \{ -24\left (\frac{M_{B_q}}{m_j+m_i}\right )^2+36\right \} 
\tilde f_6(x), \nonumber \\
&A_4^q=\left \{ -132\left (\frac{M_{B_q}}{m_j+m_i}\right )^2 \right \} 
 xf_6(x),\quad A_5^q=\left \{ -144\left (\frac{M_{B_q}}{m_j+m_i}\right )^2-84\right \}\tilde f_6(x).
\label{Adef}
\end{align}
Here,  we use $x=m_{\tilde g}^2/m_{\tilde q}^2$, where
$m_{\tilde g}$ is the gluino mass.
For the cases of $q=d$ and  $q=s$, we take $(i, j)= (1,3)$ and
 $(i, j)= (2,3)$, respectively, where $m_1=m_d$, $m_2=m_s$ and $m_3=m_b$.
The loop functions
$f_6(x)$ and  $\tilde f_6(x)$ are shown  in \cite{Hayakawa:2012ua}.

For the case of $x\simeq 1$,  we get $A_2\simeq -1$,
$A_3^q\simeq 30$, $A_4^q\simeq -10$ and   $A_5^q\simeq 10$.
Therefore, each term at the r.h.s. of Eq.(\ref{bbbarmixing})
  may contribute to $M_{12}^{q,SUSY}$ comparably.
However, magnitudes of 
$(\delta _d^{LR})_{ij}$  and $(\delta_d^{RL})_{ij}$ are constrained
severely by the  $b\to s\gamma $ decay as discussed later.

The squark flavor mixing can be tested in
 the CP-violating  asymmetries  in the neutral $B$ meson decays. 
Let us present the framework of these calculations.
The effective Hamiltonian for $\Delta B=1$ 
process is defined as 
\begin{equation}
H_{eff}=\frac{4G_F}{\sqrt{2}}\left [\sum _{q'=u,c}V_{q'b}V_{q's}^*\sum _{i=1,2}C_iO_i^{(q')}-V_{tb}V_{ts}^*
\sum _{i=3-6,7\gamma ,8G}\left (C_iO_i+\widetilde C_i\widetilde O_i\right )\right ],
\end{equation}
where the local operators are given as 
\begin{align}
&O_1^{(q')}=(\bar s_\alpha\gamma _\mu P_Lq_\beta')
(\bar q_\beta'\gamma ^\mu P_Lb_\alpha),
\qquad O_2^{(q')}=(\bar s_\alpha\gamma _\mu P_Lq_\alpha')
(\bar q_\beta'\gamma ^\mu P_Lb_\beta), \nonumber \\
&O_3=(\bar s_\alpha\gamma _\mu P_Lb_\alpha)\sum _q(\bar q_\beta\gamma ^\mu P_Lq_\beta),
\quad O_4=(\bar s_\alpha\gamma _\mu P_Lb_\beta)\sum _q(\bar q_\beta\gamma ^\mu P_Lq_\alpha), \nonumber \\
&O_5=(\bar s_\alpha\gamma _\mu P_Lb_\alpha)\sum _q(\bar q_\beta\gamma ^\mu P_Rq_\beta),
\quad O_6=(\bar s_\alpha\gamma _\mu P_Lb_\beta)\sum _q(\bar q_\beta\gamma ^\mu P_Rq_\alpha), \nonumber \\
&O_{7\gamma }=\frac{e}{16\pi ^2}m_b\bar s_\alpha\sigma ^{\mu \nu }P_Rb_\alpha
F_{\mu \nu }, 
\qquad O_{8G}=\frac{g_s}{16\pi ^2}m_b\bar s_\alpha\sigma ^{\mu \nu }
P_RT_{\alpha\beta}^ab_\beta G_{\mu \nu }^a,
\end{align}
where 
$P_R=(1+\gamma_5)/2$, $P_L=(1-\gamma_5)/2$, and $\alpha$ and $\beta$ are color
 indices, and $q$ is taken to be $u,d,s,c$.
Here, $C_i$'s  $\widetilde C_i$'s are the Wilson coefficients, 
and $\widetilde O_i$'s are  the operators by replacing  $L(R)$ with  $R(L)$ 
in $O_i$.
In this paper, $C_i$ includes both SM contribution and gluino   one,
such as  $C_i=C_i^{\rm SM}+C_i^{\tilde g}$, where
$C_i^{\text{SM}}$ is given in Ref.~\cite{Buchalla:1995vs} and
$C_i^{\tilde g}$ is  presented  as follows \cite{Endo:2004fx}:
\begin{align}
C_3^{\tilde g}&\simeq \frac{\sqrt{2}\alpha _s^2}{4G_FV_{tb}V_{ts}^*m_{\tilde q}^2}(\delta _d^{LL})_{23}
\left [-\frac{1}{9}B_1(x)-\frac{5}{9}B_2(x)-\frac{1}{18}P_1(x)-\frac{1}{2}P_2(x)\right ], \nonumber \\
C_4^{\tilde g}&\simeq \frac{\sqrt{2}\alpha _s^2}{4G_FV_{tb}V_{ts}^*m_{\tilde q}^2}(\delta _d^{LL})_{23}
\left [-\frac{7}{3}B_1(x)+\frac{1}{3}B_2(x)+\frac{1}{6}P_1(x)+\frac{3}{2}P_2(x)\right ], \nonumber \\
C_5^{\tilde g}&\simeq \frac{\sqrt{2}\alpha _s^2}{4G_FV_{tb}V_{ts}^*m_{\tilde q}^2}(\delta _d^{LL})_{23}
\left [\frac{10}{9}B_1(x)+\frac{1}{18}B_2(x)-\frac{1}{18}P_1(x)-\frac{1}{2}P_2(x)\right ], \nonumber \\
C_6^{\tilde g}&\simeq \frac{\sqrt{2}\alpha _s^2}{4G_FV_{tb}V_{ts}^*m_{\tilde q}^2}(\delta _d^{LL})_{23}
\left [-\frac{2}{3}B_1(x)+\frac{7}{6}B_2(x)+\frac{1}{6}P_1(x)+\frac{3}{2}P_2(x)\right ], \nonumber \\
C_{7\gamma }^{\tilde g}&\simeq -\frac{\sqrt{2}\alpha _s\pi }{6G_FV_{tb}V_{ts}^*m_{\tilde q}^2}
\Bigg [(\delta _d^{LL})_{23}\left (\frac{8}{3}M_3(x)-
\mu \tan \beta \frac{m_{\tilde g}}{m_{\tilde q}^2}\frac{8}{3}M_a(x)\right )
+(\delta _d^{LR})_{23}\frac{m_{\tilde g}}{m_b}\frac{8}{3}M_1(x)\Bigg ], \nonumber \\
C_{8G}^{\tilde g}&\simeq -\frac{\sqrt{2}\alpha _s\pi }{2G_FV_{tb}V_{ts}^*m_{\tilde q}^2}
\Bigg [(\delta _d^{LL})_{23}\Bigg \{ \left (\frac{1}{3}M_3(x)+3M_4(x)\right ) \nonumber \\
&-\mu \tan \beta \frac{m_{\tilde g}}{m_{\tilde q}^2}\left (\frac{1}{3}M_a(x)+3M_b(x)\right )\Bigg \} 
+(\delta _d^{LR})_{23}\frac{m_{\tilde g}}{m_b}\left (\frac{1}{3}M_1(x)+3M_2(x)\right )\Bigg ].
\label{Coeff}
\end{align}
Here the double mass insertion is included in $C_{7\gamma }^{\tilde g}$
and $C_{8G}^{\tilde g}$.
The Wilson coefficients  $\widetilde C_i^{\tilde g}$'s are 
obtained by replacing $L(R)$ with $R(L)$ in  $C_i^{\tilde g}$'s.
The loop functions, which we use in our calculations, are presented
in our previous paper \cite{Hayakawa:2012ua}.

The CP-violating  asymmetries $\mathcal{S}_f$ in Eq.~(\ref{sf}) are 
 calculated by using $\lambda_f$, which is given 
 for  $B_d^0\to \phi K_S$ and $B_d^0\to \eta 'K^0$ as follows:
\begin{align}
\lambda_{\phi K_S,\  \eta 'K^0}&=-e^{-i\phi _d}\frac{\displaystyle \sum _{i=3-6,7\gamma ,8G}
\left (C_i^\text{SM}\langle O_i \rangle+C_i^{\tilde g}\langle O_i \rangle+
\widetilde C_i^{\tilde g}\langle \widetilde O_i \rangle \right )}
{\displaystyle \sum _{i=3-6,7\gamma ,8G}
\left (C_i^{\text{SM}*}\langle O_i \rangle+C_i^{{\tilde g}*}
\langle O_i \rangle+\widetilde C_i^{{\tilde g}*}\langle\widetilde O_i \rangle \right )}~,  
\label{asymBd}
\end{align}
where  $\langle O_i \rangle$ is the abbreviation of
  $\langle f |O_i | B_q^0\rangle$.
It is noticed that  
$\langle\phi K_S|O_i|B_d^0\rangle=\langle\phi K_S|
\widetilde O_i|B_d^0\rangle $
and $\langle\eta' K^0|O_i|B_d^0\rangle=-\langle\eta' K^0|
\widetilde O_i|B_d^0\rangle$
because of  the parity of the final state.
We  have also $\lambda_{f}$ for  $B_s^0\to\phi \phi$
as follow:
\begin{align}
\lambda_{\phi \phi}&=e^{-i\phi _s}\frac{\displaystyle \sum _{i=3-6,7\gamma ,8G}
C_i^\text{SM}\langle O_i \rangle+C_i^{\tilde g}\langle O_i \rangle+
\widetilde C_i^{\tilde g}\langle \widetilde O_i \rangle}
{\displaystyle \sum _{i=3-6,7\gamma ,8G}
C_i^{\text{SM}*}\langle O_i \rangle+C_i^{{\tilde g}*}
\langle O_i \rangle+\widetilde C_i^{{\tilde g}*}\langle\widetilde O_i \rangle}~, 
\label{asymBs}
\end{align}
with 
$\langle\phi\phi|O_i|B_s^0\rangle=-\langle\phi\phi|\widetilde O_i|B_s^0\rangle $.

In these  non-leptonic decays, the $C_{8G}^{\tilde g}\langle O_{8G}\rangle$
dominates these amplitude, but 
small  contributions from other  Wilson coefficients
 are also  taken account in  our calculations.
Therefore, we estimate each  hadronic matrix elements
by using the factorization relations in Ref.~\cite{Harnik:2002vs}. 

Let us discuss the each contribution of the mass insertion parameters
 to  $C_{8G}^{\tilde g}$ in Eq.(\ref{Coeff}).
Taking account that the loop functions $M_i(x)$ are of  same order
and $m_{\tilde g}\simeq m_{\tilde q}$, the ratio of 
$LL$ component and $LR$ one is
$(\delta_d^{LL})_{23}\times \mu\tan\beta/m_{\tilde q}$ to
$(\delta_d^{LR})_{23}\times m_{\tilde q}/m_b$.
If  ${\cal O}(\mu\tan\beta)\simeq {\cal O}(m_{\tilde q})$ and
$m_{\tilde q}\geq 1$ TeV, the $LR$ component may  contribute  significantly
to $C_{8G}^{\tilde g}$  due to the enhancement factor  
$m_{\tilde q}/m_b={\cal O}(10^2)$.
For example, in the case of $(\delta_d^{LL})_{23}=10^{-2}$
and  $(\delta_d^{LR})_{23}=10^{-3}$, the $LR$ component
dominate $C_{8G}^{\tilde g}$, while it is minor 
in $M_{12}^{q,\text{SUSY}}$ as seen in  Eq.(\ref{bbbarmixing}).
This situation is also kept in the  $b\to s\gamma $ decay.

The $b\to s\gamma $ decay is a typical one to investigate new physics.
The branching ratio is given as
\begin{equation}
\frac{BR(B \to X_s \gamma )}
{BR(B \to X_c  e \bar{\nu_e})}
=
\frac{|V_{ts}^*V_{tb}|^2}
{|V_{cb}|^2}
\frac{6 \alpha}{\pi f(z)}
|C_{7\gamma}^{\text{eff}}|^2 ,
\end{equation}
where
\begin{equation}
f(z)
=
1-8z+8z^3-z^4-12z^2 \text{ln}z \ ,  \qquad 
z = \frac{m_{c}^2} {m_{b}^2}.
\end{equation}
Here, $C_{7\gamma}^{\text{eff}}$ includes both contributions
from the SM and  the gluino-squark flavor mixing $C_{7\gamma}^{\tilde g}$.
As seen  in Eq.(\ref{Coeff}), both  $C_{7\gamma}^{\tilde g}$ and 
  $C_{8G}^{\tilde g}$ have the  similar dependence of
  $(\delta _d^{LR})_{23}$.
Therefore, we should discuss carefully the contribution
  from $(\delta _d^{LR})_{23}$ in our numerical calculations.

We can discuss the direct CP violation  $A_{\text{CP}}^{b\to s\gamma}$
in the $b\to s\gamma $ decay, which  is given as \cite{Kagan:1998bh}
\begin{equation}
\begin{split}
A_{\text{CP}}^{b \to s \gamma} 
&=
\left.
\frac{\Gamma(\bar{B} \to X_s \gamma) - \Gamma(B \to X_{\bar{s}} \gamma)}
{\Gamma(\bar{B} \to X_s \gamma) + \Gamma(B \to X_{\bar{s}} \gamma)}
\right|_{E_{\gamma} > (1-\delta) E_{\gamma}^{\text{max}}} \\
&=
\frac{\alpha_s (m_b)}
{|C_{7\gamma}|^2}
\Big [
\frac{40}{81}
\text{Im} \small[ C_2 C_{7\gamma}^* \small]
-
\frac{8 z}{9} \small[v(z)+b(z, \delta)\small ]
\text{Im}\Big[\left (1+\frac{V_{us}^* V_{ub}}{V_{ts}^* V_{tb}}\right  )
C_2 C_{7\gamma}^*\Big] \nonumber \\
&
-\frac{4}{9}
\text{Im} \small[ C_{8G} C_{7\gamma}^* \small]
+
\frac{8z}{27}
b(z,\delta) \text{Im} \Big[ \left(1+\frac{V_{us}^* V_{ub}}{V_{ts}^* V_{tb}} \right ) C_2 C_{8G}^* \Big]
\Big ],
\end{split}
\end{equation}
where $v(z)$ and $b(z,\delta)$ are explicity given in  \cite{Kagan:1998bh},
and $C_i^{\text{eff}}$ includes both the SM and SUSY contributions.
Although the experimental data has still large error bar,
 we can discuss  the SUSY contribution to 
 $A_{\text{CP}}^{b\to s\gamma}$.


Let us set up the framework of our calculations.
Suppose that $\mu\tan\beta$ is at most ${\cal} O(1)$TeV.
Then, magnitudes of $(\delta _d^{LL})_{23}$ and $(\delta _d^{RR})_{23}$
are constrained by $M_{12}^s$ as seen in Eq.(\ref{bbbarmixing}).
Taking account of $h_s=0.1$ in Eq.(\ref{hsigmaLHCb}),
  we obtain  $|(\delta _d^{LL})_{23}|\simeq |(\delta _d^{RR})_{23}|\simeq 0.02$
 in our  previous work \cite{Hayakawa:2012ua}.
Then,   these contributions to  $C_{7\gamma}^{\tilde g}$ and 
$C_{8G}^{\tilde g}$ are minor.

On the other hand,  $(\delta _d^{LR})_{23}$ and $(\delta _d^{RL})_{23}$
are severely  constrained by  $C_{7\gamma}^{\text{eff}}$ and 
$C_{8G}^{\text{eff}}$.
We show the constraint for   
$(\delta _d^{LR})_{23}$ and $(\delta _d^{RL})_{23}$ 
in our following calculations.
In our convenience,  we suppose 
$|(\delta _d^{LR})_{23}|=|(\delta _d^{RL})_{23}|$.
Then, we can   parametrize these parameters  as follows:
\begin{equation}
(\delta _d^{LR})_{23}=|(\delta _d^{LR})_{23}|e^{2i\theta _{23}^{LR}},
\qquad (\delta _d^{RL})_{23}=|(\delta _d^{LR})_{23}|e^{2i\theta _{23}^{RL}},
\label{MILR}
\end{equation}
where $\theta _{23}^{LR}$ and $\theta _{23}^{RL}$ are taken
 in the region $[0-\pi]$.
By using this set up, we show  numerical analyses in the next section. 

\section{Numerical analyses}
\label{sec:Numerical}

In this section, we show the numerical analyses of the CP-violation
 in the $B$ mesons. 
In our following numerical calculations,
 we fix the squark mass and the gluino mass as
\begin{equation}
m_{\tilde q}=1000~\text{GeV},\qquad m_{\tilde g}=1500~\text{GeV},
\end{equation}
which are consistent with recent lower bound of these masses at LHC
\cite{Aad:2011ib}.
We use  relevant parameters as given in \cite{Hayakawa:2012ua}
to estimate  the SM contribution.

At first, we discuss  the $b\to s\gamma$ decay.
The observed $b\to s\gamma$ branching ratio is 
$(3.60\pm  0.23)\times 10^{-4}$ \cite{PDG}, on the other hand 
the SM prediction is given as $(3.15\pm  0.23)\times 10^{-4}$
at ${\cal{O}}(\alpha_s^2)$
\cite{Buras:1998raa,Misiak:2006zs}.
Since $\mu\tan\beta$ is supposed to be lower than ${\cal O}(1~{\rm TeV})$,
the contribution of $(\delta _d^{LL})_{23} \simeq (\delta _d^{RR})_{23}$
 is negligibly small. 
The  contribution of $(\delta _d^{LR})_{23}$ becomes  important
 through the interference with the SM component
in the decay amplitude. On the other hand,  since $(\delta _d^{RL})_{23}$ 
does not interfere with the SM component, its contribution is minor.
 The branching ratio gives the constraint for the magnitude 
 of $(\delta _d^{LR})_{23}$. 
The direct CP violation of the $b\to s\gamma$ is also
useful to constraint  $(\delta _d^{LR})_{23}$. 
We show the $A_{\text{CP}}^{b \to s \gamma}$
versus $|(\delta _d^{LR})_{23}|$ in Figure 1,
where the upper and lower bounds of the experimental data with $90\%$ C.L.
 are denoted red lines, and the predicted value of the SM 
is  shown  by the green line as the  eye guide.
 As far as $|(\delta _d^{LR})_{23}|\leq 10^{-3}$,
 the predicted value is within the experimental allowed region.

In Figure 2, 
we show the  $|(\delta _d^{LR})_{23}|$ dependence of the branching ratio
taking accont   the constraint of $A_{\text{CP}}^{b \to s \gamma}$
as seen in Figure 1.
 Here,  the allowed  region at $|(\delta _d^{LR})_{23}|=0$
 is the SM prediction.
As the magnitude of    $(\delta _d^{LR})_{23}$ increases,
 the predicted  region of the branching ratio splits into the larger
 region and smaller one.
The excluded region between two regions   is due to 
the constraint of $A_{\text{CP}}^{b \to s \gamma}$.
Then,  
 the predicted branching ratio becomes inconsistent with
 the experimental data at $|(\delta _d^{LR})_{23}|\geq 5.5\times 10^{-3}$.

\begin{figure}[b!]
\begin{minipage}[]{0.45\linewidth}
\vspace{4 mm}
\includegraphics[width=8cm]{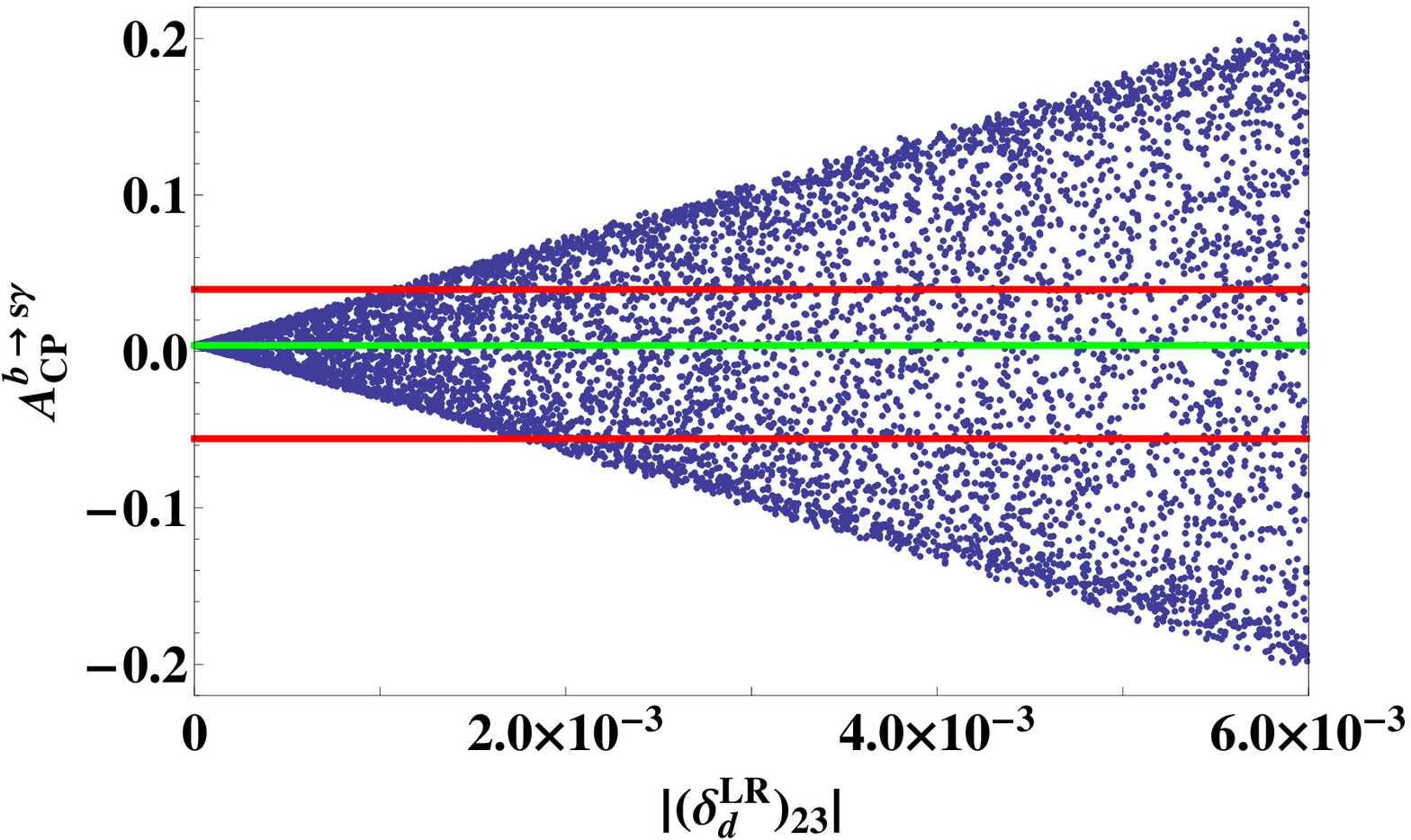}
\caption{The direct CP violation  $A_{\text{CP}}^{b \to s \gamma}$
versus $|(\delta _d^{LR})_{23}|$,
where the green line denotes the SM prediction
and  red lines denote the upper and lower bounds of the experimental data with $90\%$ C.L..}
\end{minipage}
\hspace{5mm}
\begin{minipage}[]{0.45\linewidth}
\includegraphics[width=8cm]{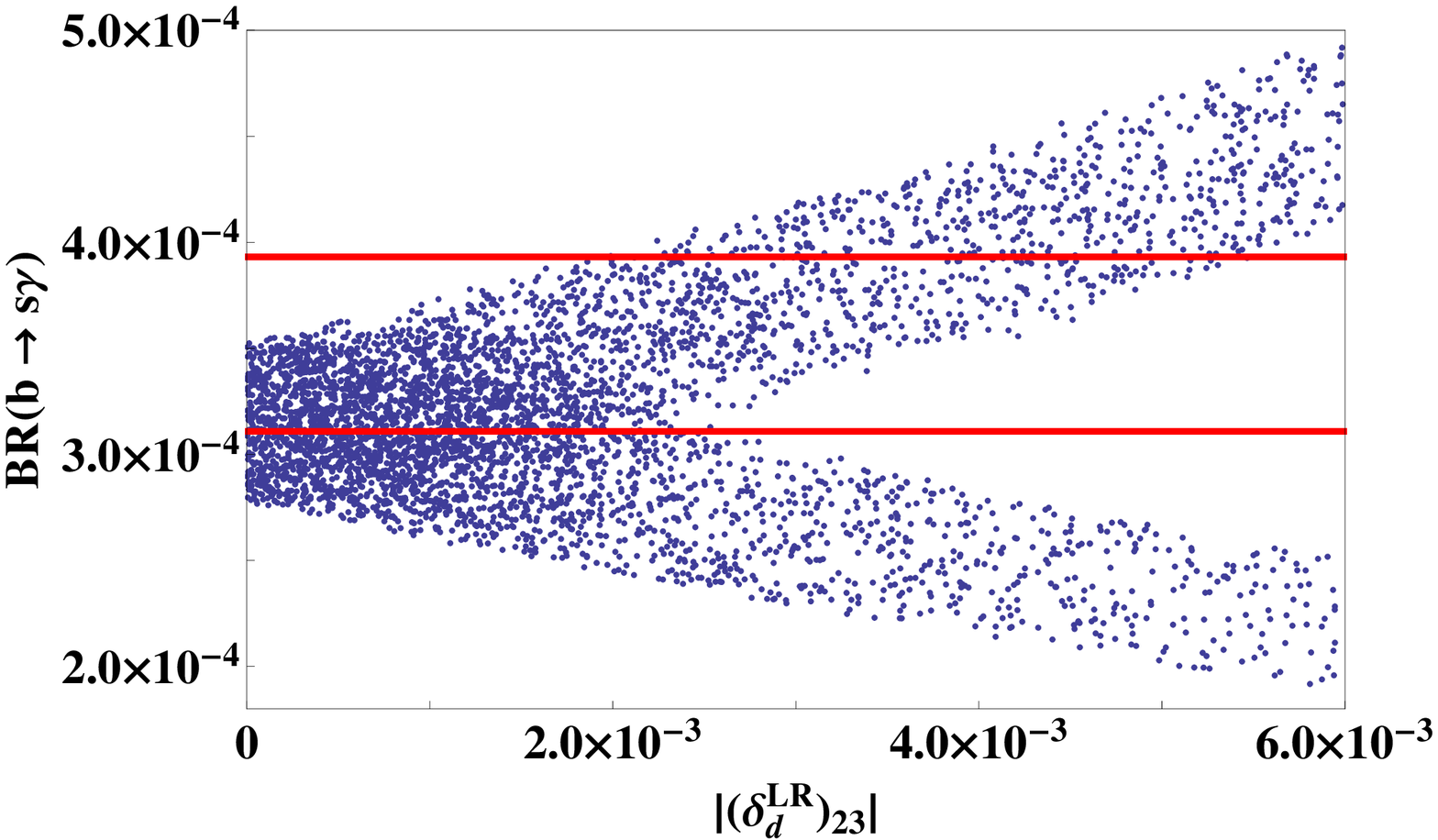}
\caption{The predicted branching ratio of  $b\to s\gamma$ 
versus $|(\delta _d^{LR})_{23}|$, where the experimental constraint
of $A_{\text{CP}}^{b \to s \gamma}$ is taken account.
Predicted value at $|(\delta _d^{LR})_{23}|=0$ is the SM one.}
\end{minipage}
\end{figure}

In order to see the role of the  phase  $\theta _{23}^{LR}$, 
 we show  $A_{\text{CP}}^{b \to s \gamma}$
versus  $\theta _{23}^{LR}$ for $|(\delta _d^{LR})_{23}|= 10^{-3}$(blue)
and $|(\delta _d^{LR})_{23}|= 10^{-4}$(orange) in Figure 3.
 The pink horizontal lines denote
the  experimental upper and lower bounds at  $1\sigma$ level.
As seen in this figure,
we find  that
 the reduction  of the experimental error-bar will constrain the SUSY
phase  $\theta _{23}^{LR}$ severely.

In Figure 4,
we plot the allowed region 
on the   $|(\delta _d^{LR})_{23}|-\theta _{23}^{LR}$ plane
by putting  the experimental data at $90\%$ C.L. of
the branching ratio and the direct CP violation
$A_{\text{CP}}^{b \to s \gamma}$.
 The allowed region of  $|(\delta _d^{LR})_{23}|$ is cut 
at $5.5\times 10^{-3}$,
where   $\theta _{23}^{LR}$ is tuned around $\pi/2$.
Around $\pi/4$ and  $3\pi/4$,  $A_{\text{CP}}^{b \to s \gamma}$
give the severe constraint as seen in Figure 3.
This  CP violation phase
  also contributes on the CP-violating asymmetry
 of the non-leptonic decays of  $B_d^0$ and $B_s^0$ mesons.

Let us discuss ${\mathcal S}_{f}$, which is the measure
 of the CP-violating asymmetry, 
for  $B_d^0\to {J/\psi  K_S}, \ {\phi K_S}, \  {\eta' K^0}$.
As discussed in Section 2, 
these   ${\mathcal S}_{f}$'s are predicted
 to be same ones in the SM.
On the other hand, 
 if the squark  flavor mixing contributes to the decay process 
 at the one-loop level, these asymmetries are different
 from among as seen in  Eq.(\ref{asymBd}).

\begin{figure}[b!]
\begin{minipage}[]{0.45\linewidth}
\includegraphics[width=8cm]{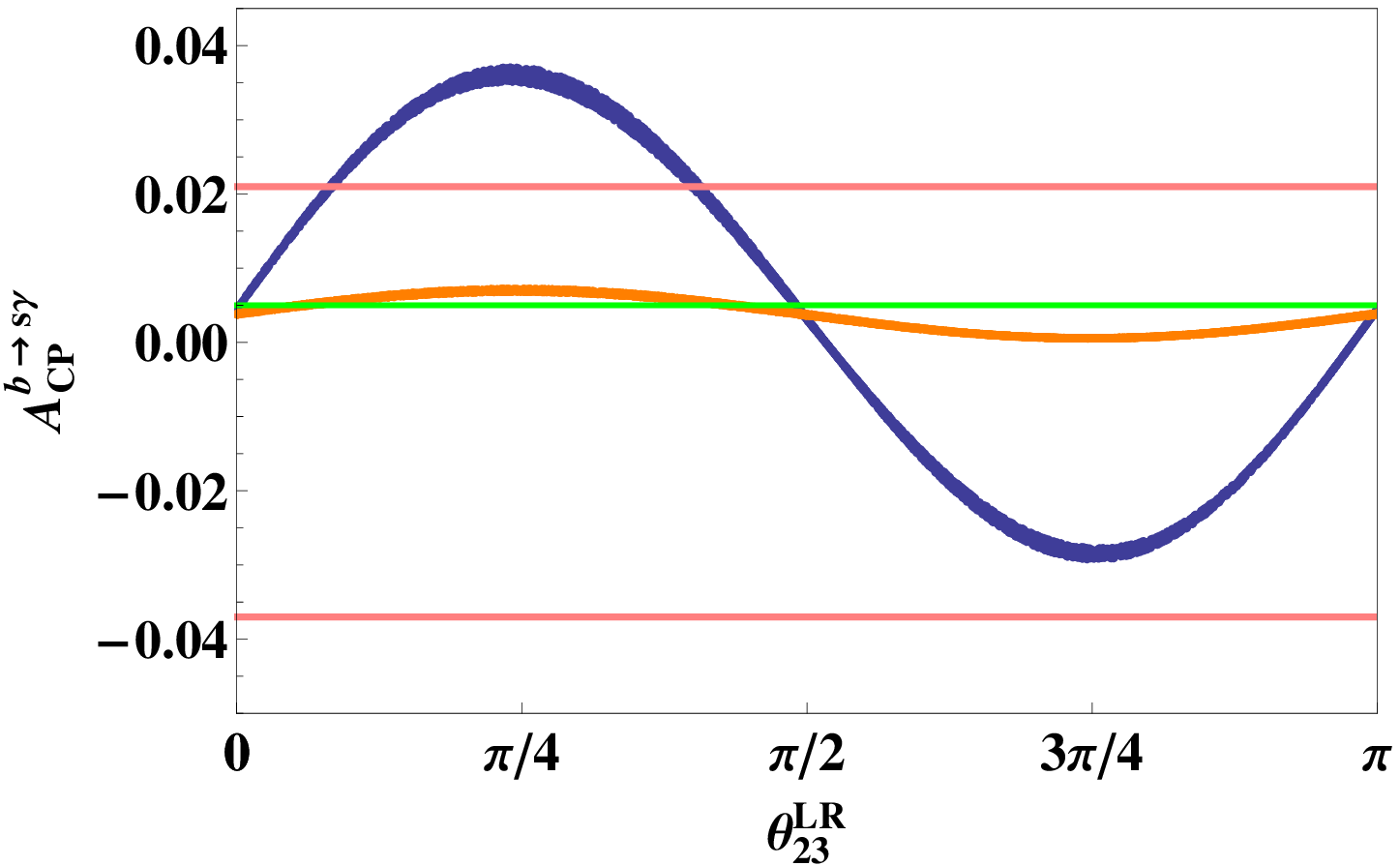}
\caption{$A_{\text{CP}}^{b \to s \gamma}$
versus  $\theta _{23}^{LR}$ for $|(\delta _d^{LR})_{23}|= 10^{-3}$(blue)
and $10^{-4}$(orange), where
the pink  lines denote
the  experimental upper and lower bounds at  $1\sigma$ level.}
\end{minipage}
\hspace{5mm}
\begin{minipage}[]{0.45\linewidth}
\vspace{0mm}
\includegraphics[width=8cm]{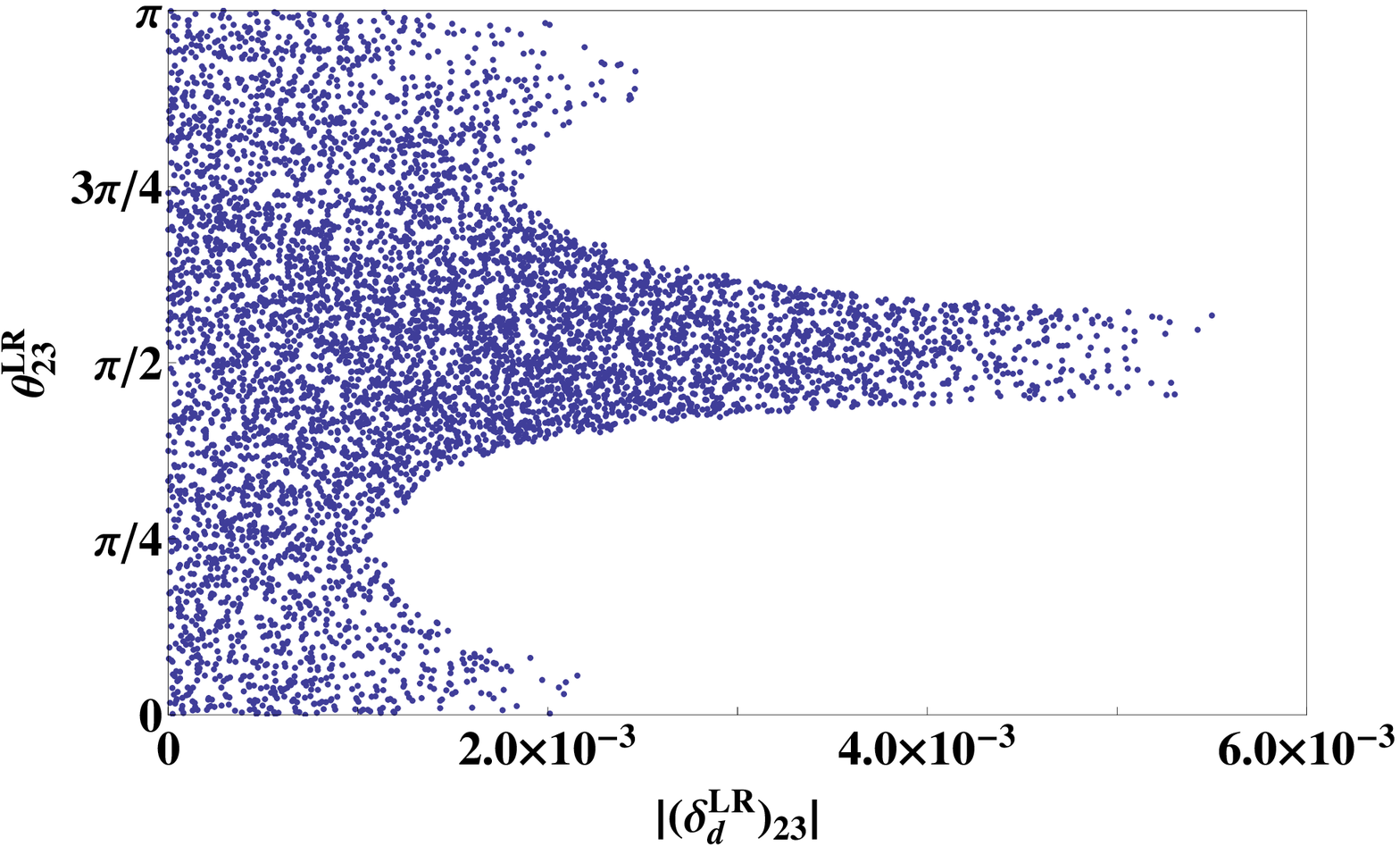}
\caption{The allowed region
 on  $|(\delta _d^{LR})_{23}|$-  $\theta _{23}^{LR}$ plane.
The experimental data at $90\%$ C.L. of
the branching ratio and 
$A_{\text{CP}}^{b \to s \gamma}$ are taken account.}
\end{minipage}
\end{figure}

\begin{figure}[b!]
\begin{minipage}[]{0.45\linewidth}
\includegraphics[width=8cm]{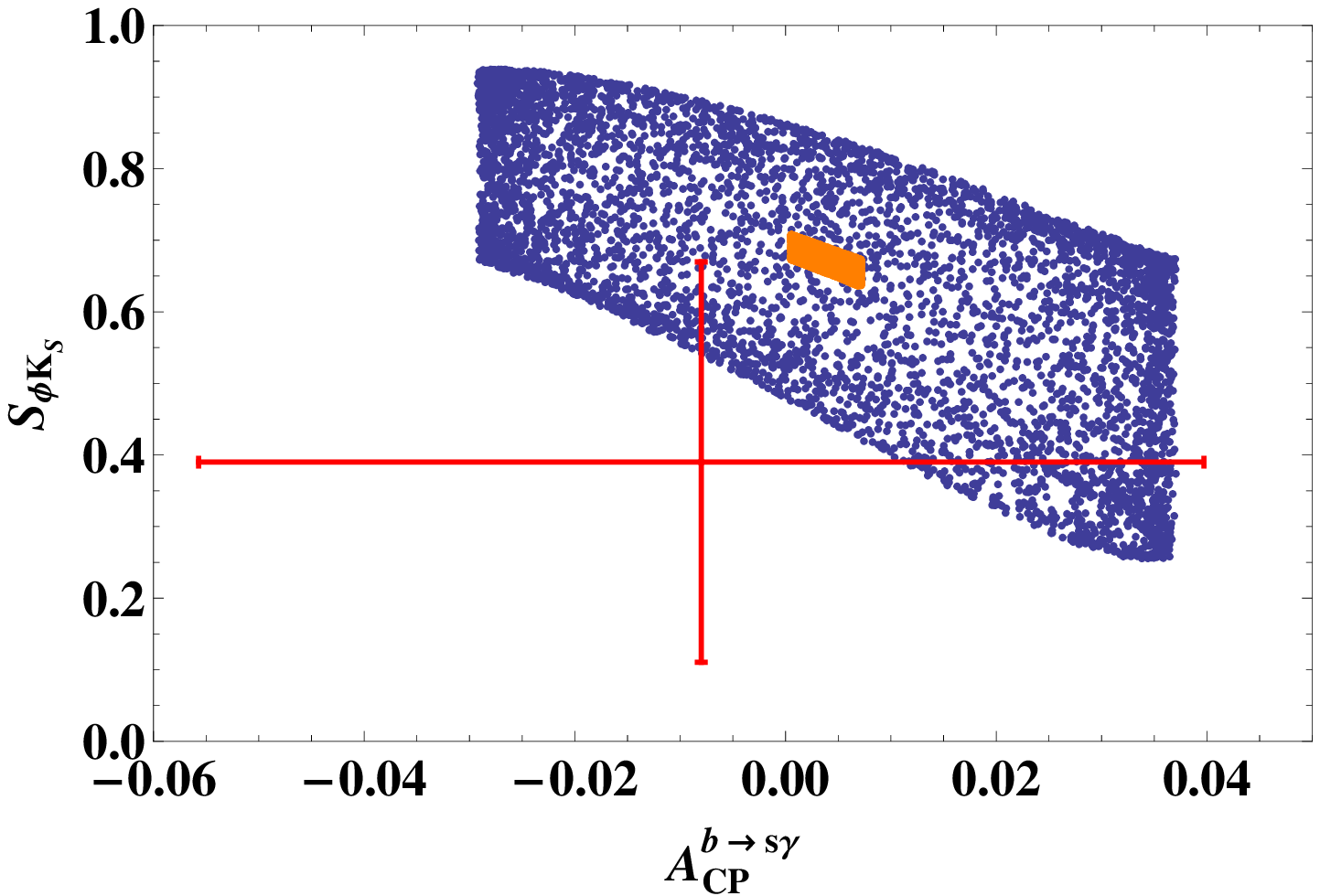}
\caption{$S_{\phi K_S}$ of $B_d^0$ versus 
 $A_{\text{CP}}^{b \to s \gamma}$ for $|(\delta _d^{LR})_{23}|= 10^{-3}$(blue)
and $10^{-4}$(orange).}
\end{minipage}
\hspace{5mm}
\begin{minipage}[]{0.45\linewidth}
\vspace{-4mm}
\includegraphics[width=8cm]{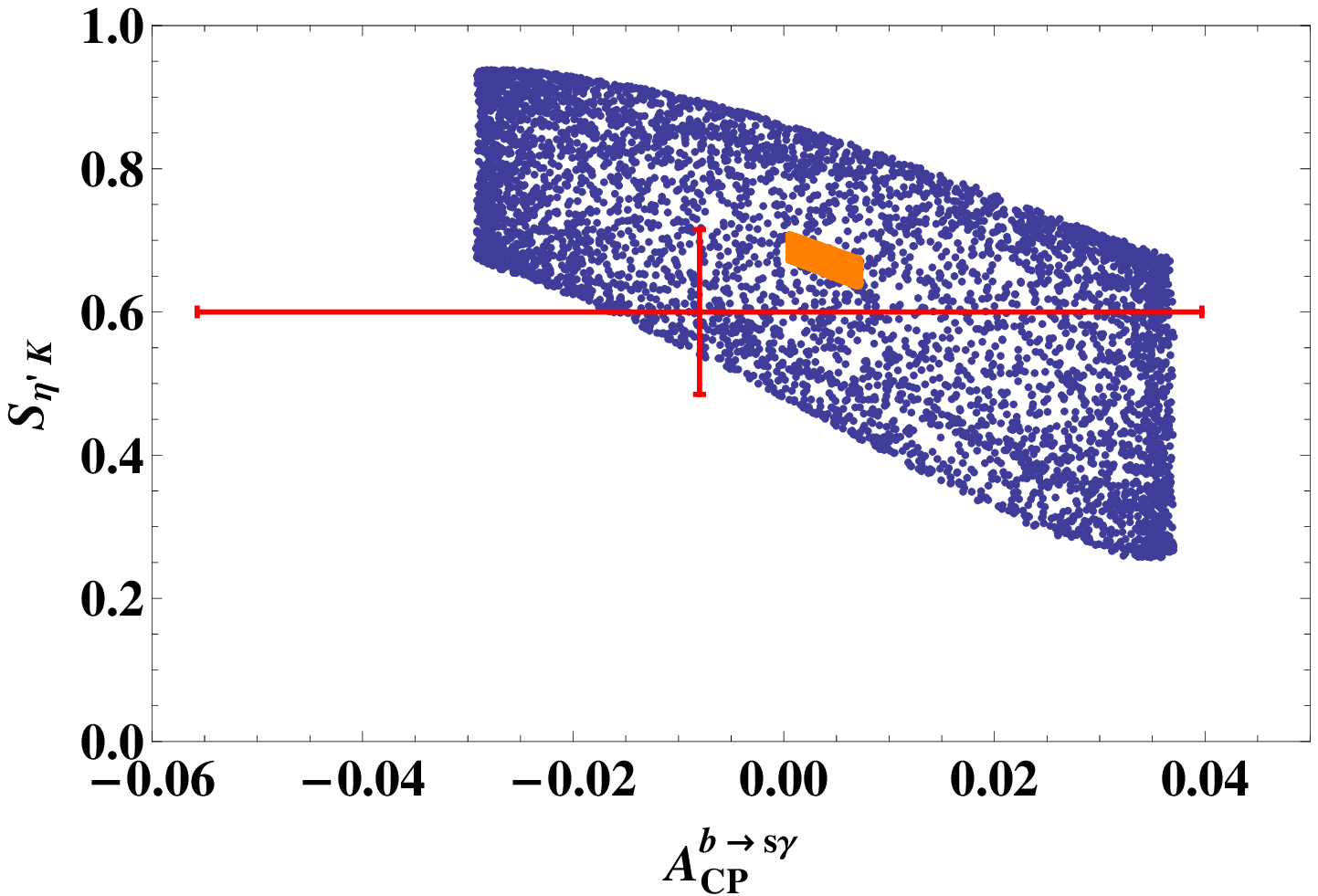}
\caption{$S_{\eta 'K^0}$ of $B_d^0$ versus 
 $A_{\text{CP}}^{b \to s \gamma}$ for $|(\delta _d^{LR})_{23}|= 10^{-3}$(blue)
and $10^{-4}$(orange).}
\end{minipage}
\end{figure}

Since  the  phase $\theta _{23}^{LR}$  contributes to 
 $A_{\text{CP}}^{b \to s \gamma}$,  
$\mathcal{S}_{\phi K_S}$ and $\mathcal{S}_{\eta 'K^0}$ of $B_d^0$ decays.
We expect the correlations among them.
We  fix  $|(\delta _d^{LR})_{23}|=10^{-4}$(orange)
and $10^{-3}$(blue) for typical values  in the following calculations.
we show the predicted regions on the  
$A_{\text{CP}}^{b \to s \gamma}$-$\mathcal{S}_{\phi K_S}$ and
 $A_{\text{CP}}^{b \to s \gamma}$-$\mathcal{S}_{\eta 'K^0}$ planes
 in Figures 5 and 6, respectively.
The experimental data is denoted by red lines at $90\%$ C.L..
We also present 
the predicted region on
 the $\mathcal{S}_{\phi K_S}$-$\mathcal{S}_{\eta 'K^0}$ plane in Figure 7,
the slant dashed line denotes the SM prediction 
$\mathcal{S}_{J/\psi K_S}=\mathcal{S}_{\phi K_S}=\mathcal{S}_{\eta 'K}$,
where the observed value
 $\mathcal{S}_{J/\psi K_S}=0.671\pm 0.023$ is put.
The reduction of the experimental error of
 $A_{\text{CP}}^{b \to s \gamma}$ will give us severe predictions for 
$\mathcal{S}_{\phi K_S}$ and $\mathcal{S}_{\eta 'K^0}$.
It is noticed that this predicted region   is different 
from the one in the previous work
\cite{Hayakawa:2012ua}, where $(\delta _d^{LR})_{23}$ is neglected
and  $\mu\tan\beta={\cal O}(10~{\rm TeV})$.

As seen in Figures 5, 6 and 7,
 the reduction of the experimental errors will provide powerful tool
to find the contribution of the squark flavor mixing.
At last, we show 
the correlation between $A_{\text{CP}}^{b \to s \gamma}$ and  
$\mathcal{S}_{\phi \phi}$ of the $B_s^0$ decay in Figure 8.
We expect the observation of the CP violation in
  $B_s^0\to \phi \phi$ at LHCb.

\begin{figure}[h!]
\begin{minipage}[]{0.45\linewidth}
\includegraphics[width=8cm]{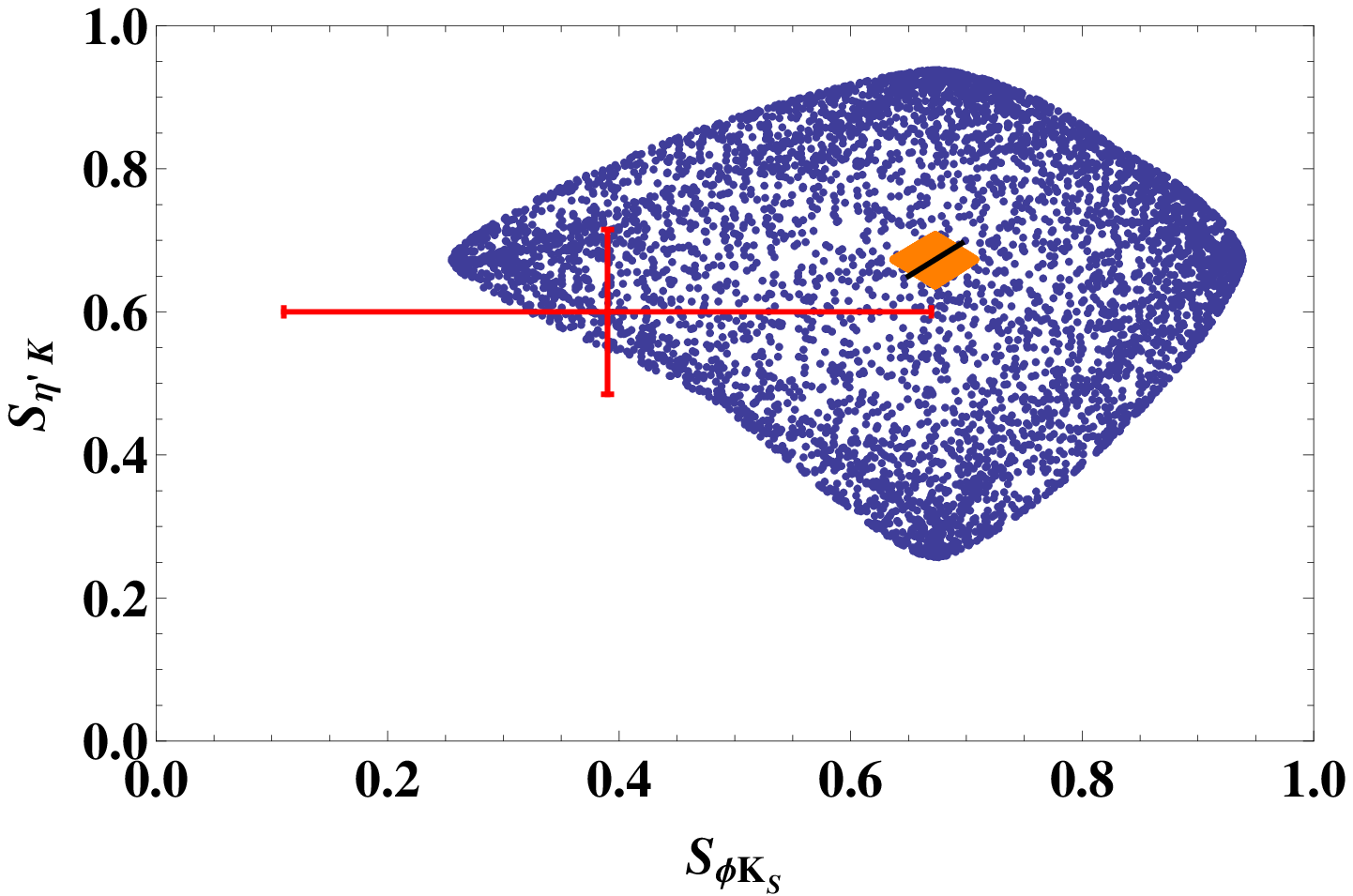}
\caption{Predeicted region on the $S_{\phi K_S}$-$S_{\eta 'K^0}$ plane, where
the slant dashed line denotes the SM prediction.}
\end{minipage}
\hspace{5mm}
\begin{minipage}[]{0.45\linewidth}
\vspace{-4mm}
\includegraphics[width=8cm]{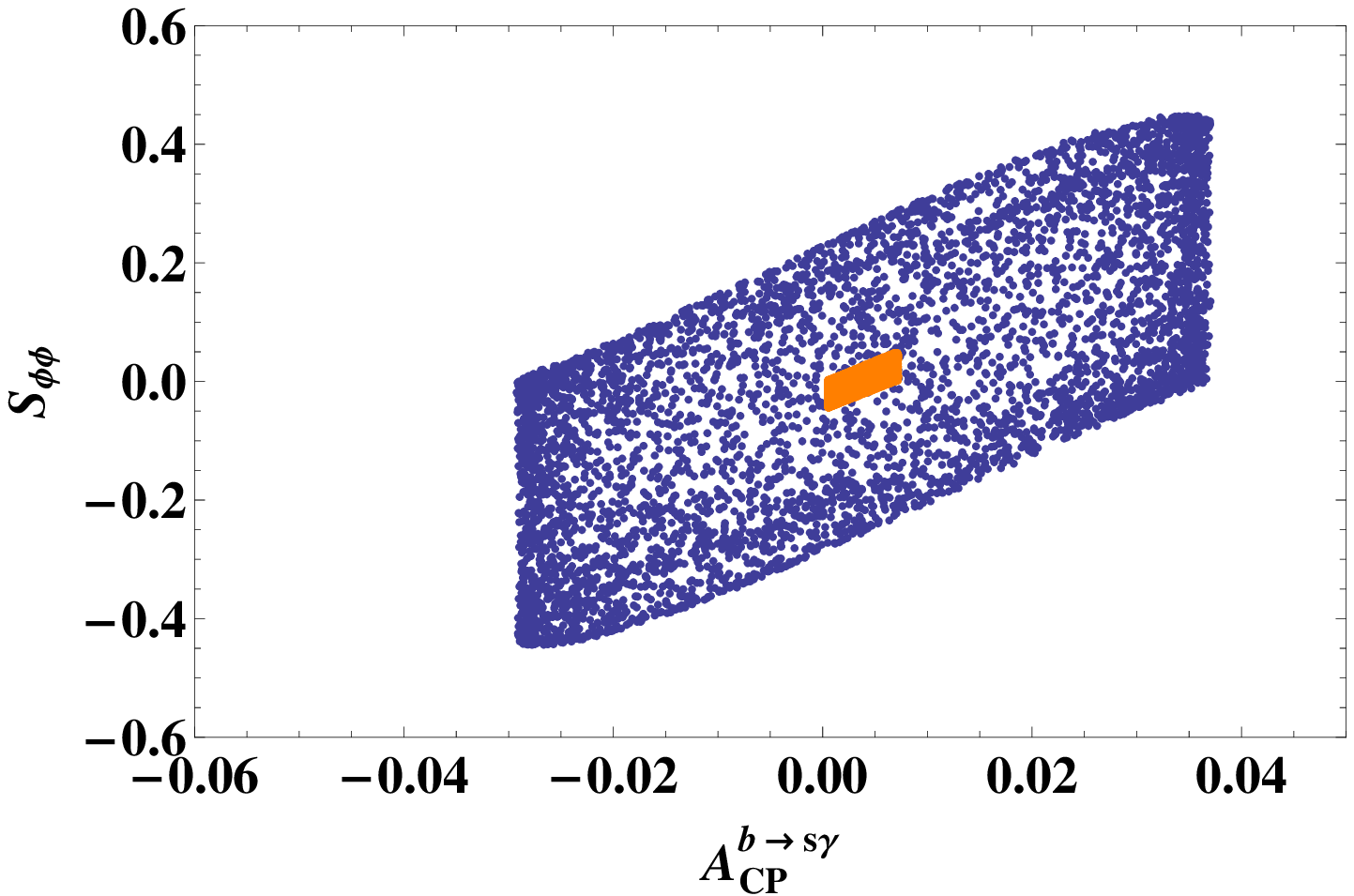}
\caption{Predicted  asymmetry  $S_{\phi \phi }$ in the $B_s^0$ decay
versus 
 $A_{\text{CP}}^{b \to s \gamma}$.}
\end{minipage}
\end{figure}


\section{Summary}
\label{sec:Summary}
The CP violation of the neutral $B$ meson is the important phenomenon
to search for  new physics. 
We have discussed the contribution of the squark flavor mixing 
from  $(\delta _d^{LR})_{23}$ and $(\delta _d^{RL})_{23}$
on the direct CP violation of the $b\to s\gamma$ decay and
the CP-violating asymmetry in the non-leptonic decays of 
$B_d^0$ and $B_s^0$ mesons.

The magnitude of $|(\delta _d^{LR})_{23}|$ 
is bounded by the branching ratio of  $b\to s\gamma$ 
 with  the constraint of $A_{\text{CP}}^{b \to s \gamma}$.
 The predicted branching ratio becomes inconsistent with
 the experimental data at $|(\delta _d^{LR})_{23}|\geq 5.5\times 10^{-3}$.
We have obtained  the allowed region 
on the   $|(\delta _d^{LR})_{23}|$-$\theta _{23}^{LR}$ plane.
 While the $|(\delta _d^{LR})_{23}|$ is cut at $5.5\times 10^{-3}$,
CP-violating phase  $\theta _{23}^{LR}$ is severely constrained
 at  $|(\delta _d^{LR})_{23}|\geq 2\times 10^{-3}$.
This  CP-violating phase also contribute 
to the CP-violating asymmetry in the non-leptonic decays of 
$B_d^0$ and $B_s^0$ mesons.

We have predicted the correlation among $A_{\text{CP}}^{b \to s \gamma}$ and  
$\mathcal{S}_{f}$ of the $B_d^0$ and $B_s^0$ decays.
These CP-violating asymmetries could deviate from
the SM predictions.

Since we suppose 
rather small $\mu\tan\beta, \ {\cal O}(1~{\rm TeV})$,
the contribution from $LL(RR)$ components are minor
 in these CP-violating asymmetries.
In this case,  the new physics contribution
 is minor in $M_{12}^q$ of $B_q-\bar B_q \ (q=d,s)$ mixing
since $|(\delta _d^{LR})_{23}|$ is at most ${\cal O}(10^{-3})$.
This result is consistent with the recent result
 of the CP violations at  LHCb as discussed in Section 2.

 In the near future,
 the precise data of 
the direct CP violation of  $b\to s\gamma$ and  CP-violating asymmetries
 in the non-leptonic decays of 
$B_d^0$ and $B_s^0$ mesons  give us the crucial  test
 for  our  framework of the squark flavor mixing.

\vspace{0.5 cm}
\noindent
{\bf Acknowledgement}

We thank Drs. E. Ko, Y. Okada, and T. Morozumi for useful discussions. 
We also thank A. Hayakawa for his help.
M.T. is  supported by JSPS Grand-in-Aid for Scientific Research,
21340055 and 24654062.


\end{document}